\documentclass[aps,pre,preprint,unsortedaddress,floats,showpacs]{revtex4-1}
\usepackage{amsmath,amssymb}
\usepackage{graphicx}
\usepackage{color}

\newcommand {\apgt} {\ {\raise-.5ex\hbox{$\buildrel>\over\sim$}}\ }
\newcommand {\aplt} {\ {\raise-.5ex\hbox{$\buildrel<\over\sim$}}\ }

\begin{document}

\title{Comparison of dust charging between Orbital-Motion-Limited theory and Particle-In-Cell simulations}

\author{Gian Luca Delzanno}
\email{delzanno@lanl.gov}
\author{Xian-Zhu~Tang}
\email{xtang@lanl.gov}

\affiliation{Theoretical Division, Los Alamos National Laboratory, Los
  Alamos, NM 87545}

\date{\today}

\begin{abstract}


The Orbital-Motion-Limited (OML) theory has been modified to predict
the dust charge and the results were contrasted with the Whipple
approximation [Tang and Delzanno, \textit{Phys. Plasmas} {\bf 21},
  123708 (2014)].  To further establish its regime of applicability,
in this paper the OML predictions (for a non-electron-emitting, spherical dust grain at rest in a collisionless, unmagnetized plasma)
are compared with Particle-In-Cell
simulations that retain the absorption radius effect.  It is found
that for large dust grain radius $r_d$ relative to the plasma Debye length $\lambda_D$,
the revised OML theory remains a very good approximation as, for the
parameters considered ($r_d/\lambda_D\le10$, equal electron and ion
temperatures), it yields the dust charge to
within $20\%$ accuracy. This is a substantial improvement over the Whipple
approximation. The dust collected currents and energy fluxes, which remain the same in the revised and standard OML theories,
are accurate to within $15-30\%$.

\end{abstract}

\maketitle

\section{Introduction}

The Orbital-Motion-Limited (OML) theory
\cite{mott-smith-langmuir-pr-1926,alpert-book-1965,laframboise-thesis-1966}
is the most widely used dust charging theory.  In its simplest form
(i.e. without dust electron-emission processes), it neglects barriers
to ion current collection that are created by local maxima of the
effective potential (the so-called absorption radius
concept~\cite{allen-etal-jpp-2000}).  This approximation results in
simple expressions for the collected currents and energy fluxes to the
grain, and allows to calculate the steady-state dust potential without
solving Poisson's equation.

Until recently OML was known to be unable to calculate the plasma
response to the charging process, that is, the OML Poisson equation could not be solved even numerically.  
This was pointed out by Allen \textit{et al.} \cite{allen-etal-jpp-2000}, who used the OML plasma
densities calculated in Ref.  \cite{alpert-book-1965} to show that OML
is intrinsically inconsistent. In particular, they showed that OML is valid when the shielding potential
decreases more slowly than $1/r^2$, but that in a Maxwellian plasma this condition
is always violated in the quasi-neutral plasma region away from the dust grain
(for any dust grain size, at least if the electron temperature is greater or equal than
the ion temperature). Allen \textit{et al.} \cite{allen-etal-jpp-2000} attributed
this inconsistency to the absorption radius effect, which is missing in OML since the effective potential
is assumed to be monotonic.  

These limitations imply that OML could not self-consistently calculate the dust charge, 
a quantity of fundamental importance for dust transport
studies since it controls the magnitude of the electromagnetic force: 
the dust charge depends on how the charged dust is screened by the plasma, namely on how 
the plasma rearranges around the dust grain in response to the charging process, and requires the full solution of Poisson's equation.
In practice, this has restricted the successful application
of OML to cases where the dust radius is much smaller than the plasma
Debye length, where one can accurately obtain the dust charge from the
dust potential by either neglecting screening effects or assuming that
these effects can be described by a Debye-Huckel potential near the
grain with a given screening length.  

In order to assess in very simple terms the importance of screening effects, one can start from Gauss' law relating the dust
charge $Q_d$ to the radial electric field at the grain surface $E_r$
(spherical symmetry is assumed where $r$ is the radial coordinate):
\begin{equation}
Q_d=4 \pi \varepsilon_0 r_d^2 \left.E_r\right|_{r=r_d}
\label{gauss}
\end{equation}
($\varepsilon_0$ is vacuum permittivity and $r_d$ is the dust radius).
A standard Debye shielding calculation allows one to express $E_r=-d \phi/dr$ in terms of the Debye-Huckel electrostatic potential $\phi$
given by 
\begin{equation}
\phi=\phi_d \frac{r_d}{r} \exp\left(-\frac{r-r_d}{\lambda_{lin}}\right),
\label{dh}
\end{equation}
where $\phi_d=\phi(r_d)$ is the dust potential and the screening length corresponds to the linearized Debye length
\begin{equation}
\frac{1}{\lambda_{lin}^2}=\frac{1}{\lambda_{De}^2}+\frac{1}{\lambda_{Di}^2},
\label{llin}
\end{equation}
with the electron (ion) Debye length given by
$\lambda_{De,i}=\sqrt{\varepsilon_0 T_{e,i}/e^2/n_\infty}$ (subscripts
'e' and 'i' label electrons and ions, respectively, $e$ is the
elementary charge, $T_{e,i}$ are the electron and ion temperatures
expressed in eV, $n_\infty$ is the unperturbed plasma density and we
have assumed singly charged ions). It follows that
\begin{equation}
Q_d=4 \pi \varepsilon_0 r_d \left(1+\frac{r_d}{\lambda_{lin}} \right)\phi_d,
\label{whipple}
\end{equation}
which is referred to as the Whipple approximation \cite{whipple-rpp-1981}. 
Obviously, in the limit $r_d/\lambda_{lin} \ll 1$ Eq. (\ref{whipple}) simply recovers
Coulomb's law applied to a charged dust particle in vacuum,
\begin{equation}
Q_d=4 \pi \varepsilon_0 r_d \phi_d,
\label{coulomb}
\end{equation}
where one recognizes the familiar dust capacitance $C_d=4 \pi \varepsilon_0 r_d$.
Equation (\ref{whipple}) indicates that when the dust grain radius becomes comparable to or larger than the screening length,
plasma screening effects cannot be neglected and might in fact be dominant in the calculation of the dust charge. One should also keep in mind
that in a fully self-consistent calculation both the dust potential and the screening length will depend on the dust radius, making the
relation between the dust charge and the dust radius even more non-linear \cite{daugherty92}.

While the small dust-radius-to-Debye-length limit works well for many
dusty plasma applications
\cite{shukla-book-2001,fortov-book-2010,bouchoule-book-1999},
important emerging applications can be in the opposite limit. 
For instance, this is the case of magnetic fusion energy applications~\cite{krash11}, where the size of dust particles emitted from the tokamak walls can easily 
be comparable to or larger than the local Debye length
\cite{sharpe02,rudakov09}.  This is also true for dust particles
injected on purpose~\cite{rudakov09,ratynskaia13,litnovsky13,ratynskaia14b,tang-delzanno-jfe-2007}. As we will show in this
paper, neglecting screening effects in this regime can lead to an
order of magnitude underestimation of the dust charge. This could have important consequences for dust particles
moving in the magnetized sheath near the tokamak walls, where the dynamics perpendicular to the wall is governed by the balance between
electrostatic and drag forces \cite{krash04,delzanno-tang-pop-2014} and for a negatively charged particles a stronger 
electrostatic force would mean a higher probability of escape, wider excursions away from the wall and stronger heat fluxes \cite{delzanno-tang-pop-2014}.
In general, however, it is worth pointing out that it appears
that most of the dust transport codes for magnetic fusion
energy applications~\cite{pigarov05,martin_epl08,bacharis10,smirnov-etal-ppcf-2007,ratynskaia13,ratynskaia14} (which are all based on OML)
use Eq. (\ref{coulomb}) to relate the dust charge to the dust
potential, i.e. neglect screening effects.

Recently Tang and Delzanno \cite{tang_pop14} showed that the problems
with the classic OML theory arise from a mistake in the integration
limits used in the derivation of the ion density. Upon correcting this
error, a revised OML theory that can now calculate the plasma response and, consequently,
the dust charge \cite{bystrenko02,tang_pop14}. This theoretical development opens up the possibility to perform a
true comparison between the OML theory and a Particle-In-Cell
(PIC) approach that calculates the charging of a dust grain in a
plasma from first principles (i.e. without invoking the OML
approximation), and assess the validity/accuracy of OML in the large
dust-radius-to-Debye-length limit.  This is the goal of the present
paper. 

We note that this study is conducted with the most basic form of OML and it is the first necessary
step towards properly accounting for screening effects.
Other effects such as the presence of an ion flow, electron emission processes from the dust grain and magnetic field
effects on current collection can be very important in magnetic fusion energy applications and their impact on dust charging 
and screening will have to be evaluated. Some of these effects do break the spherical symmetry on which OML rests,
and require an upgraded theory together with fully self-consistent simulations. Whether some form of the OML theory that accounts for
these non-OML effects in a simplified but reasonable way can still be formulated is, at this point, an open question.

The paper is organized as follows. In Sec. \ref{PIC} we briefly
describe the PIC simulations and the revised OML theory.  In
Sec. \ref{results} we present a comparison between the two in terms of
average dust charge, potential, collected currents and energy fluxes,
and the plasma densities and electrostatic potential. In
Sec. \ref{concl} we draw the conclusions of this study.

\section{Particle-In-Cell simulations and revised OML theory} \label{PIC}
\subsection{Particle-In-Cell simulations}

In this subsection we describe the PIC simulations of the charging and
shielding of a dust grain in a plasma whose steady state is compared
with that obtained by the revised OML theory.

The PIC simulations are conducted in spherical geometry and in the
collisionless, electrostatic limit (i.e. the plasma particles move
only in response to the self-consistent electric field generated in
the system), with the same code used for instance in Refs.
\cite{delzanno04,delzanno05,delzanno-etal-ieee-2013,delzanno14}.  The spherical
dust grain of radius $r_d$ is at rest in the center of the system and
its surface corresponds to the inner boundary of the simulation
domain. The outer boundary is a concentric sphere of radius $R$. At
the outer boundary the system is open and we inject a Maxwellian
plasma at every time step, to compensate for the plasma particles that
leave the system due to their thermal motion. Moreover, the
electrostatic potential is set to zero at the outer boundary.
Particles that hit the dust grain are removed from the simulation and
their charge is accumulated on the dust grain. This translates into a
boundary condition on the electric field given by Gauss' law (we
assume a perfectly conducting dust grain).

At time $t=0$ we load a Maxwellian plasma with temperature $T_e=T_i$,
density $n_e=n_i=n_\infty$ and the ion-to-electron mass ratio
$m_i/m_e$ ($\lambda_D=\lambda_{De}=\lambda_{Di}$). We choose $m_i/m_e=1836$. 
For the other simulation
parameters, we use a grid with uniform radial spacing and $N_r=1000$
grid points, while the time step is $\omega_{pe}\Delta t=0.05$
[$\omega_{pe}=\sqrt{e^2 n_\infty/\left(\varepsilon_0 m_e \right)}$ is
  the electron plasma frequency].  The electron thermal velocity is
defined as $v_{th,e}=\sqrt{T_e/m_e}$.  We conduct a parametric study
of dust charging varying the dust radius in the range
$r_d/\lambda_D=0.25-10$. The outer domain boundary is at $R/\lambda_D=30$
for all the runs except when $r_d/\lambda_D=10$, where we use $R/\lambda_D=60$
(and $N_r=2000$ to keep the same level of resolution).
The final time of the simulations is
$\omega_{pe}T\sim2000$ for $r_d/\lambda_D \ge1$ and
$\omega_{pe}T\sim4000$ for $r_d/\lambda_D <1$.  At the end of the
simulations the average number of particles per cell is $\sim
3500-4000$ for $r_d/\lambda_D \ge1$ and $\sim 7000$ for $r_d/\lambda_D
<1$.

Initially the electrons charge the grain negatively because of their higher mobility. This creates a sheath electric
field that eventually equilibrates the electron ($I_e$) and ion ($I_i$) currents to the grain and a steady state is
reached when $I_e+I_i=0$ (floating condition) with the grain negatively charged.
In the results presented in the next section, the steady-state data from PIC simulations
is averaged over the last quarter of the simulations (labeled as $T_{ave}$), corresponding to $\omega_{pe}T_{ave}\sim 500$ for $r_d/\lambda_D \ge1$
and $\omega_{pe}T_{ave}\sim 1000$ for $r_d/\lambda_D <1$.

\subsection{Revised OML theory}
The steady state calculated from the revised OML theory amounts to solving Poisson's equation in spherical geometry
\begin{equation}
\nabla^2\phi = \frac{e}{\varepsilon_0}\left(n_e^{OML} -n_i^{OML}\right), \label{oml-poisson}
\end{equation}
with the following expressions for the ion density~\cite{bystrenko02,tang_pop14}
\begin{eqnarray}
 \frac{n_i^{OML}(z)}{n_{\infty}}
&= &
\sqrt{-\frac{\beta\varphi}{\pi}}
  \left[1 + \sqrt{1 - \frac{\varphi_d^{OML}}{z^2\varphi}}\right] 
+ \frac{e^{-\beta\varphi}}{2} 
\left[
1 - {\rm Erf}\left(\sqrt{-\beta\varphi}\right)\right] + \nonumber \\
&&  \frac{\sqrt{1- z^{-2}}}{2}
e^{- \beta\tilde{\varphi}} 
\left[ 1 - {\rm Erf}\left(\sqrt{-\beta\tilde{\varphi}}\right) \right],\,\,\,\,\,\,\,\,\, \phi(z)<\frac{\phi_d^{OML}}{z^2}; \label{ni-oml} \\
\frac{n_i^{OML}(z)}{n_{\infty}}
&=&
 \sqrt{-\frac{\beta\varphi}{\pi}}
+ \frac{e^{- \beta\varphi}}{2}
\left[
1 - {\rm Erf}\left(\sqrt{- \beta\varphi}\right)\right]
+ \frac{\sqrt{1- z^{-2}}}{2}
  e^{- \beta\tilde{\varphi}},\,\,\,\,\,\,\,\,\, \phi(z)\ge\frac{\phi_d^{OML}}{z^2},
  \label{ni-oml-new}
\end{eqnarray}
and electron density
\begin{eqnarray}
\frac{n_e^{OML}(z)}{n_{\infty}} &=&
\frac{1}{2}\left\{
1 + {\rm Erf}\left(\sqrt{\varphi-\varphi_d^{OML}}\right)
+ \sqrt{1-z^{-2}}\left[1 - {\rm Erf}\left(\sqrt{\frac{\varphi-\varphi_d^{OML}}{1-z^{-2}}}\right)
\right]
\exp\left[\frac{\varphi-\varphi_d^{OML}}{z^2-1}\right]
\right\} \nonumber \\
&& \exp \left(\varphi\right).
\label{ne-oml}
\end{eqnarray}
In Eqs. (\ref{ni-oml}), (\ref{ni-oml-new}) and (\ref{ne-oml}) we have defined the following quantities:
$z= r/r_d$, $\varphi= e\phi/T_e$, $\beta= T_e/T_i$, $\tilde{\varphi} = \left(\varphi -\varphi_d^{OML}/z^2\right)/(1-z^{-2})$
and the dust potential is $\phi_d^{OML}=\phi(r_d)$.
Poisson's equation (\ref{oml-poisson}) is solved with the following boundary conditions: 
at the outer boundary we have
\begin{equation}
\left.\frac{d \phi}{dr}\right|_{r=R}=-\frac{2}{R} \phi(R),
\label{outer}
\end{equation}
which mimics the asymptotic behavior of the shielding potential $\phi\sim {\rm const}/r^2$ [and can be easily verified in the quasi-neutral region by
Taylor expanding Eqs. (\ref{ni-oml-new}) and (\ref{ne-oml}) in the limit $z\gg1$ and $\varphi\ll1$], while
the floating potential condition is used on the dust grain
\begin{equation}
I_e(\phi_d^{OML})+I_i(\phi_d^{OML})=0,
\label{floatcond}
\end{equation}
with currents given by
\begin{eqnarray}
  I_i & =  & e 4\pi r_d^2 n_{\infty} \sqrt{\frac{T_i}{2\pi m_i}}
  \left(1-\frac{e\phi_d^{OML}}{T_i}\right), \label{Iioml} \\
  I_e & = & - e 4\pi r_d^2 n_{\infty} \sqrt{\frac{ T_e}{2\pi m_e}} \exp\left(\frac{e\phi_d^{OML}}{T_e}\right) \label{Ieoml}.
\end{eqnarray}
The dust charge is obtained by the solution of Poisson's equation via Gauss' law (\ref{gauss}).
Furthermore, the OML power collected by the grain from the plasma is obtained by \cite{ticos06}
\begin{eqnarray}
&& q_i= T_i \left(\displaystyle{\frac{2 -\displaystyle{\frac{e \phi_d^{OML}}{T_i}}}{1 -\displaystyle{\frac{e \phi_d^{OML}}{T_i}}}}-\frac{e \phi_d^{OML}}{T_i} \right) \frac{I_{i}}{e} \label{qioml} \\
&& q_{e}=2 T_{e} \frac{|I_{e}|}{e}. \label{eqoml}
\end{eqnarray}

While in the remainder of the text we will generally refer to the revised OML theory for the comparison with PIC, 
we note that the floating condition (\ref{floatcond}) in the revised OML theory has not changed relative to the standard OML theory
and therefore the two theories produce the same value of floating potential, dust currents and energy fluxes. 
We reiterate that in the revised OML theory one can now solve Poisson's equation
to get the screening near the grain and calculate the dust charge, which in general could not be done in the standard OML theory because of the inconsistencies discussed
by Allen \textit{et al.} \cite{allen-etal-jpp-2000}.

For the discussion of the results of the next section, Eq. (\ref{oml-poisson}) has been solved numerically with a second-order accurate finite difference scheme.
We introduce a relative error defined as
\begin{equation}
\delta \varepsilon_f=\frac{|f^{PIC}-f^{OML}|}{|f^{PIC}|},
\label{error}
\end{equation}
where $f$ represents a generic quantity of interest. Here and in the remainder of the paper we will use superscripts 'PIC' and 'OML' where necessary to indicate the values of $f$
obtained from PIC simulations or from the revised OML theory.

\section{Results and discussion} \label{results}

Table \ref{t1} shows results concerning the dust charge $Q_d$, the dust potential $\phi_d$ and the parameter
\begin{equation}
\Gamma=\frac{Q_d}{4 \pi \varepsilon_0 r_d \phi_d}-1,
\label{gamma}
\end{equation}
obtained from the PIC simulations and the revised OML theory. As expected, the dust charge increases with the dust radius. 
A least-squares fit of the data between $r_d/\lambda_D=3$ and $r_d/\lambda_D=10$ shows that $Q_d\propto r_d^{1.6}$,
indicating that the non-linear corrections to Eq. (\ref{coulomb}), i.e. the screening from the plasma, is important.
 Moreover, there is a good agreement between the PIC simulations and the revised OML theory: up to $r_d/\lambda_D\sim 5$, 
 $\delta \varepsilon_Q<10\%$, and the biggest discrepancy obtained at $r_d/\lambda_D=10$ is only $\delta \varepsilon_Q\simeq20\%$.
 Most importantly, using the Whipple approximation (\ref{whipple}) overestimates the dust charge by a factor of $2.5$:
 for $r_d/\lambda_D=10$, $Q_d/\left(e n_\infty \lambda_D^3 \right)\simeq-4757$ ($\lambda_{lin}/\lambda_D=1/\sqrt{2}$).
 If one had used expression (\ref{coulomb})
 (which is appropriate only when screening effects can be neglected), the dust charge would be grossly underestimated:
 for $r_d/\lambda_D=10$, $Q_d/\left(e n_\infty \lambda_D^3 \right)\simeq-314$, a factor of $6$ smaller than that obtained by PIC.
 Similar considerations can be drawn by looking at the dust potential. While the OML dust potential is independent of the dust radius,
 $\displaystyle{\frac{e\phi^{OML}_d}{T_e}}=-2.5$, the PIC simulations show that the dust potential becomes more negative as the dust
 radius increases. These results are consistent with those of Ref. \cite{willis-etal-psst-2010}, and reflect the fact that as the dust radius grows, current
 collection becomes similar to that occurring in planar geometry rather than in spherical geometry (where plasma particles might miss the 
 dust grain because of angular momentum effects and the related centrifugal force). The agreement between PIC and revised OML is
 very good up to  $r_d/\lambda_D\sim10$, with $\delta \varepsilon_\phi < 12\%$. 

\begin{table}
\centering
\caption{Summary of the PIC and revised OML simulations: dust charge and potential. The dust potential obtained from OML is 
$\displaystyle{\frac{e\phi^{OML}_d}{T_e}}=-2.5$.}
\begin{tabular}{|c||c|c||c||c|c|}
\hline
\hline
$\displaystyle{\frac{r_d}{\lambda_D}}$ & $\displaystyle{\frac{Q_d^{PIC}}{e n_\infty \lambda_D^3}}$ 
& $\displaystyle{\frac{Q_d^{OML}}{e n_\infty \lambda_D^3}}$ & $\displaystyle{\frac{e\phi^{PIC}_d}{T_e}}$ & $\Gamma^{PIC}$ & $\Gamma^{OML}$\\
\hline
$0.25$ & $-10.00$ & $-9.87$ & $-2.51$ & $0.27$ & $0.25$\\
$0.5$ & $-23.23$ & $-23.03$ & $-2.51$ & $0.47$& $0.46$\\
$1$ & $-57.64$ & $-57.46$ & $-2.54$ &  $0.81$ & $0.83$\\
$2$ & $-152.02$ & $-153.97$ & $-2.57$ &  $1.36$ & $1.45$\\
$3$ & $-277.39$ & $-286.24$ & $-2.60$ &  $1.83$ & $2.03$\\
$4$ & $-431.49$ & $-455.00$ & $-2.64$ &  $2.25$ & $2.62$\\
$5$ & $-612.52$ & $-661.41$ & $-2.68$ &  $2.63$ & $3.20$\\
$10$ &  $-1898.75$ & $-2282.87$ & $-2.82$ & $4.36$ & $6.26$\\
\hline
\hline
\end{tabular}
\label{t1}
\end{table}

Table \ref{t2} shows the electron and ion currents collected by the dust grain at steady state, again varying the dust radius.
Note that Table \ref{t2} only reports $I_e^{PIC}$ since at steady state the dust grain is at floating potential
(although the floating condition is only satisfied in an average sense, $I_e^{PIC}+I_i^{PIC}\simeq0$). 
One can see that the electron current, normalized to a reference current $I_{norm}=e n_\infty v_{th,e} r_d^2$, decreases for larger dust
radii, consistent with the fact that the dust potential is becoming more negative:
for $r_d/\lambda_D=10$, $I_e/I_{norm}$ is about $27\%$ less than that for $r_d/\lambda_D \aplt 1$.
Furthermore, Table \ref{t2} also presents the ratio of the collected plasma currents to those
obtained from the OML theory using however the dust potential obtained by PIC simulations (Table \ref{t1}) instead of $e \phi_d^{OML}/T_e=-2.5$.
One can see that taking into account the change of the dust potential with $r_d$ results in an electron collected current in excellent agreement with expression (\ref{Ieoml}). 
For the ions, on the other hand, this is true only for $r_d/\lambda_D \aplt 1$ and there is a progressively larger departure
from expression (\ref{Iioml}) as $r_d$ grows: for $r_d/\lambda_D=10$, $I_i^{PIC}/I_i^{OML}(\phi_d^{PIC})=0.68$.

\begin{table}
\centering
\caption{Summary of the PIC simulations: dust currents. For the PIC simulations, the ion current is equal to the electron current since at steady state
the floating condition $I_e^{PIC}+I_i^{PIC}\simeq0$ is satisfied. Note that for the fourth and fifth columns 
the OML currents are calculated from expressions (\ref{Iioml}) and (\ref{Ieoml}) but using the dust potential obtained from PIC (and shown in Table \ref{t1}) instead of
$e \phi_d^{OML}/T_e=-2.5$.}
\begin{tabular}{|c||c|c|c|c||}
\hline
\hline
$\displaystyle{\frac{r_d}{\lambda_D}}$ & $\displaystyle{\frac{I_e^{PIC}}{e n_\infty v_{th,e} r_d^2}}$ & $\displaystyle{\frac{I_e^{PIC}}{I_e^{OML}(\phi_d^{OML})}}$
& $\displaystyle{\frac{I_e^{PIC}}{I_e^{OML}(\phi_d^{PIC})}}$ & $\displaystyle{\frac{I_i^{PIC}}{I_i^{OML}(\phi_d^{PIC})}}$\\
\hline
$0.25$ & $-0.40$ & $0.97$ & $0.98$ & $0.97$ \\
$0.5$ & $-0.41$ & $1.00$ & $1.01$  & $0.99$\\
$1$ & $-0.40$ & $0.96$ & $1.00$ & $0.96$\\
$2$ & $-0.39$ & $0.94$ & $1.01$ & $0.93$\\
$3$ & $-0.37$ & $0.90$ & $1.00$ & $0.88$\\
$4$ & $-0.36$ & $0.87$  & $1.00$ & $0.84$\\
$5$ & $-0.34$ & $0.83$ & $1.00$ & $0.80$\\
$10$ &  $-0.30$ & $0.73$ & $1.00$ & $0.67$\\
\hline
\hline
\end{tabular}
\label{t2}
\end{table}

Table \ref{t3} shows the electron $q_e^{PIC}$ and ion $q_i^{PIC}$ power collected by the dust grain obtained from the PIC simulations,
with the same format of Tables \ref{t1} and \ref{t2}. As expected, the results (which are normalized to a reference power, 
$q_{norm}=T_e e n_\infty v_{th,e} r_d^2$) are consistent with those in Table \ref{t2} and show a variation of the dust collected power which
is comparable to that of the dust collected currents.

\begin{table}
\centering
\caption{Summary of the PIC simulations: power collected by the dust grain. Note that in the fourth and seventh columns 
the OML collected power is calculated using the dust potential obtained from PIC and shown in Table \ref{t1}.}
\begin{tabular}{|c||c|c|c||c|c|c|||}
\hline
\hline
$\displaystyle{\frac{r_d}{\lambda_D}}$  & $\displaystyle{\frac{q_e^{PIC}}{T_e n_\infty v_{th,e}r_d^2}}$ & $\displaystyle{\frac{q_e^{PIC}}{q_e^{OML}(\phi_d^{OML})}}$ & $\displaystyle{\frac{q_e^{PIC}}{q_e^{OML}(\phi_d^{PIC})}}$ &
$\displaystyle{\frac{q_i^{PIC}}{T_e n_\infty v_{th,e}r_d^2}}$ & $\displaystyle{\frac{q_i^{PIC}}{q_i^{OML}(\phi_d^{OML})}}$ & $\displaystyle{\frac{q_i^{PIC}}{q_i^{OML}(\phi_d^{PIC})}}$  \\
\hline
$0.25$ & $0.79$ &$0.96$ & $0.97$ & $1.49$ & $0.96$ & $0.96$\\
$0.5$ & $0.81$ &$0.99$ & $1.00$ & $1.56$ & $1.00$ & $1.00$\\
$1$ & $0.79$ &$0.96$& $1.00$ & $1.53$ & $0.99$ & $0.96$\\
$2$ & $0.78$ &$0.95$ & $1.01$ & $1.52$ & $0.98$ & $0.95$\\
$3$ & $0.74$ &$0.90$ & $1.00$ & $1.49$ & $0.96$ & $0.91$\\
$4$ & $0.72$ &$0.87$ & $1.00$ & $1.47$ & $0.95$ & $0.88$\\
$5$ & $0.69$ &$0.83$ & $1.00$ & $1.44$ & $0.93$ & $0.84$\\
$10$ & $0.60$ &$0.73$ & $1.00$ & $1.32$ & $0.85$ & $0.72$\\
\hline
\hline
\end{tabular}
\label{t3}
\end{table}

\begin{figure}
\begin{centering}
\includegraphics[scale=0.8]{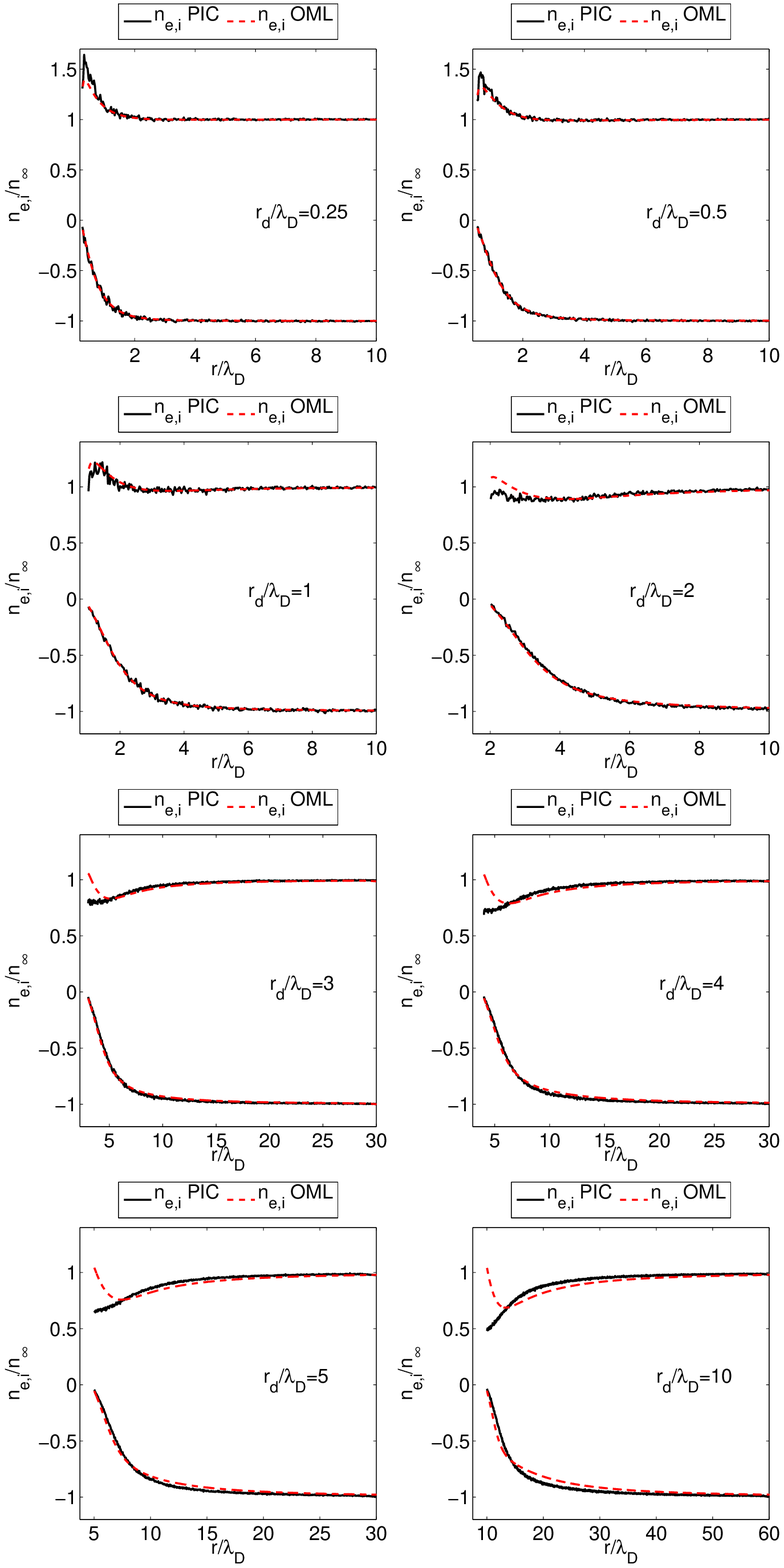}
\par\end{centering}
\caption{Comparison of the steady-state plasma densities from PIC simulations (solid lines) and from the revised OML theory (dashed lines) varying the dust radius.
For $r_d/\lambda_D\le2$ only part of the computational domain is shown.}
\label{fig:density}
\end{figure}

Figure \ref{fig:density} shows the average plasma densities obtained from PIC (solid lines) and from the revised OML theory (dashed lines).
In general, despite some standard PIC noise, there is very good agreement on the electron densities at least up to $r_d/\lambda_D \sim 5$,
with more visible differences for $r_d/\lambda_D =10$. For the ion density, on the other hand, there is reasonable agreement between theory
and simulations up to $r_d/\lambda_D \sim 1$. Although the PIC noise is still quite high for $r_d/\lambda_D \aplt 1$, some non-monotonicity in the PIC ion density might be inferred near the dust grain,
as in the revised OML theory. However, for $r_d/\lambda_D > 2$, one can see that the
PIC simulations show an ion density which increases monotonically with distance from the grain, typical of a sheath in planar geometry.
The revised OML density, however, remains monotonically decreasing near the grain, as expected from a sheath in spherical geometry \cite{daugherty92}.

\begin{figure}
\begin{centering}
\includegraphics[scale=0.8]{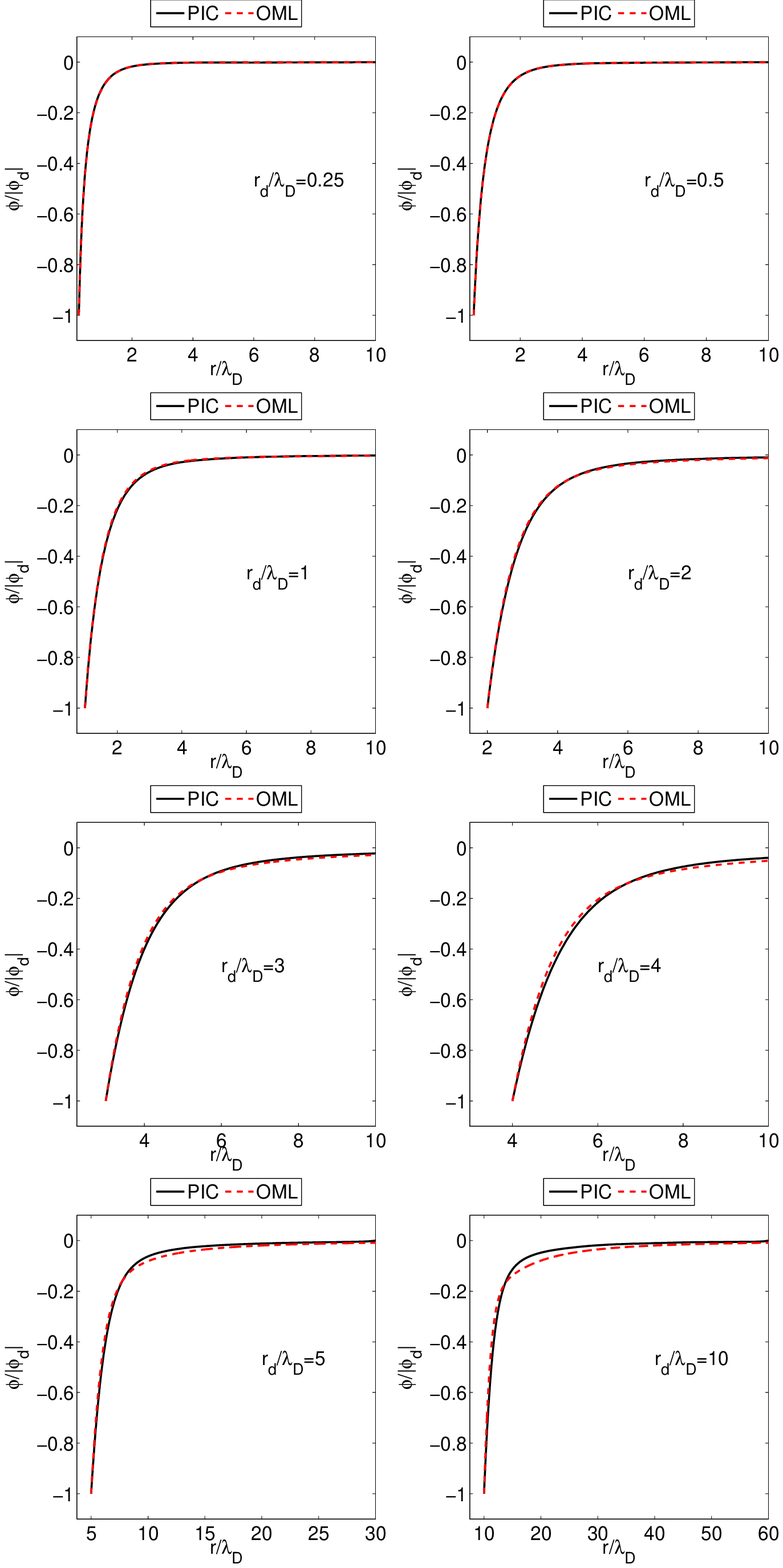}
\par\end{centering}
\caption{Comparison of the steady-state electrostatic potential from PIC simulations (solid line) and from the revised OML theory (dashed line) varying the dust radius.
For $r_d/\lambda_D\le4$ only part of the computational domain is shown.}
\label{fig:potential}
\end{figure}

Figure \ref{fig:potential} shows the average potential (normalized to the dust potential) in the simulation domain for various dust radii. 
Note that the potential for each simulation run is normalized to its respective floating potential (which depends on the dust radius, as shown in Table \ref{t1}) 
to facilitate an assessment of the screening length.
In general there is a very good
agreement between theory and simulations at least up to $r_d/\lambda_D\sim 5$ and the screening from the plasma is captured correctly by the revised OML theory
(hence the very good agreement on the dust charge already visible from Table \ref{t1}).

\section{Conclusions} \label{concl}
In this paper we have presented a comparison between PIC simulations
of the steady-state charging and shielding of a dust grain in a plasma
and the revised OML theory.  The latter was developed explicitly in
Ref.~\cite{tang_pop14} for the purpose of
evaluating the
plasma response to the charging process in the OML framework.  This
yields the dust charge including screening effects self-consistently.

We have performed a parametric study changing the dust radius in the regime of interest to magnetic fusion energy applications, i.e. when the dust grain is larger than the
plasma Debye length. We have considered the simplest case of a non-electron-emitting spherical dust grain at rest in a collisionless, unmagnetized plasma.
While the revised OML theory still cannot capture the transition from a spherical to a planar sheath that occurs when the dust becomes too large
(as shown already in Ref. \cite{willis-etal-psst-2010}), for the parameters considered (hydrogen plasma and $T_e/T_i=1$) we have shown that the revised OML theory
remains a very good approximation up to $r_d/\lambda_D\sim 10$. In particular, the dust charge is accurate to within $20\%$ once screening
effects are correctly included. For comparison, for $r_d/\lambda_D= 10$
the Whipple approximation overestimates the  dust charge by a factor of $2.5$, while neglecting screening effects completely would underestimate the dust charge by almost one order of magnitude.
Other quantities of interest to tokamak dust transport studies, like currents and energy fluxes collected by the dust from the plasma, remain the same in the revised and standard OML theories and 
are overestimated by only $15-30\%$ for $r_d/\lambda_D\aplt10$. 
The contributions to the screening of other effects that can be important in magnetic fusion energy applications, like ion-flow, electron emission and magnetic fields, will be explored in the future.

\medskip{}
\acknowledgements
This research was supported by the U.S. Department of Energy Office of
Science, Office of Fusion Energy Sciences, under the auspices of the
National Nuclear Security Administration of the U.S. Department of
Energy by Los Alamos National Laboratory, operated by Los Alamos
National Security LLC under contract DE-AC52-06NA25396.

\bibliographystyle{apsrev4-1}
\bibliography{dusty-plasma}

\pagebreak

\end{document}